\documentclass[amsmath, amssymb, reprint, aip]{revtex4-2}

\usepackage{graphicx} 
\usepackage{hyperref} 
\usepackage{mathptmx} 
\usepackage{xcolor}   
\usepackage{dcolumn}  

\DeclareMathAlphabet{\pazocal}{OMS}{zplm}{m}{n} 

\hypersetup{
    colorlinks=true,
    linkcolor=blue,
    citecolor=blue,
    filecolor=blue,
    urlcolor=blue
}

\graphicspath{{./}{./figs/}}

\newcommand{\SSD}{Sensor Science Division, National Institute of Standards and Technology, Gaithersburg, Maryland 20899, USA}
\newcommand{\QMD}{Quantum Measurement Division, National Institute of Standards and Technology, Gaithersburg, Maryland 20899, USA}
\newcommand{\JQI}{Joint Quantum Institute, College Park, Maryland 20742, USA}
\newcommand{\UMD}{Physics Department, University of Maryland, College Park, Maryland, 20742, USA}

\newcolumntype{d}{D{.}{.}{4.4}}

\begin{document}

\title{Accurate measurement of the loss rate of cold atoms due to background gas collisions for the quantum-based cold atom vacuum standard}
\author{Daniel S. Barker}
\affiliation{\SSD}
\author{James A. Fedchak}
\affiliation{\SSD}
\author{Jacek K{\l}os}
\affiliation{\JQI}\affiliation{\UMD}
\author{Julia Scherschligt}
\affiliation{\SSD}
\author{Abrar A. Sheikh}
\affiliation{\SSD}
\author{Eite Tiesinga}
\affiliation{\JQI}\affiliation{\UMD}\affiliation{\QMD}
\author{Stephen P. Eckel}
\email{stephen.eckel@nist.gov}
\affiliation{\SSD}
\date{\today}

\begin{abstract}
We present measurements of thermalized collisional rate coefficients for ultra-cold $^7$Li and $^{87}$Rb colliding with room-temperature He, Ne, N$_2$, Ar, Kr, and Xe. 
In our experiments, a combined flowmeter and dynamic expansion system, a vacuum metrology standard, is used to set a known number density for the room-temperature background gas in the vicinity of the magnetically trapped $^7$Li or $^{87}$Rb clouds.
Each collision with a background atom  or molecule removes a $^7$Li or $^{87}$Rb atom from its trap and the change in the atom loss rate with  background gas density is used to determine the thermalized loss rate coefficients with fractional standard uncertainties better than 1.6~\% for $^7$Li and 2.7~\% for $^{87}$Rb.
We find consistency---a degree of equivalence of less than one---between the measurements and recent quantum-scattering calculations of the loss rate coefficients
[J. K\l{}os and E. Tiesinga, {\it J. Chem. Phys.} {\bf 158}, 014308 (2023)], with the exception of the loss rate coefficient for both $^7$Li and $^{87}$Rb colliding with Ar.
Nevertheless, the agreement between theory and experiment for all other studied systems provides validation that a quantum-based measurement of vacuum pressure using cold atoms also serves as a primary standard for vacuum pressure, which we refer to as the cold-atom vacuum standard.
\end{abstract}

\maketitle

\section{Introduction}

Since the first magnetic trapping of laser-cooled neutral alkali-metal atoms, experiments performed in ultra-high vacuum chambers,\cite{Migdall1985} it has been recognized that collisions of residual or background gas atoms and molecules with the trapped atoms establish a limit on the lifetime of cold atoms in their shallow  magnetic trap.
Inverting the problem---using the measured loss rate of cold atoms from a conservative magnetic trap to sense vacuum pressure in the ultra-high vacuum regime---has since been pursued in several experiments.\cite{Bjorkholm1988, Fagnan2009, US8803072, Arpornthip2012, Yuan2013, Moore2015, Makhalov2016, Booth2019}
Such a conversion requires  knowledge of gas-species-dependent loss rate coefficients $L$ to determine the background-gas number densities $n$ from  measured trap loss rates $\Gamma$.
In fact, $n=\Gamma/L$ and a value for pressure $p$ then follows from the ideal gas law $p = n k T$, where $k$ is the Boltzmann constant and $T$ is the background gas temperature, assuming that this gas is in thermal equilibrium with the walls of the vacuum chamber.
The loss rate coefficients correspond to thermally averaged rate coefficients for elastic, momentum-changing collisions between a trapped atom and room-temperature background-gas atoms or molecules. 

Many of the first attempts to measure pressure with laser-cooled atoms relied on semi-classical theory of elastic scattering\cite{Child} to compute the loss rate coefficients.\cite{Bjorkholm1988, Fagnan2009, Arpornthip2012, Yuan2013, Makhalov2016}
Quantum universality of these elastic and small-angle, or diffractive, collisions, derived from this theory, has been put forward as a means to extract the loss rate coefficients.\cite{Booth2019, Shen2020, Shen2021} 
The accuracy of the semi-classical model, however, is not well characterized. 
Analyses by Refs.~\onlinecite{Shen2022, Klos2023}, for example, have suggested that loss rate coefficients based on this theory can be in error by as much as 30~\%.

Here, we measure loss rate coefficients with high-accuracy in a model-independent way.
Our measurements achieve one-standard-deviation combined statistical and systematic ($k=1$) relative uncertainties better than 1.6~\% for $^7$Li and 2.7~\% for $^{87}$Rb .
We use two different cold atom vacuum sensors\cite{Scherschligt2017, Eckel2018, Ehinger2022, Barker2022} that trap a relatively small number of either $^7$Li or $^{87}$Rb sensor atoms in a weak magnetic trap with energy depth $W$, typically $W/k\lesssim 1$~mK, connected to a dynamic expansion system, which sets a known number density of background atoms or molecules.
We compare our findings to recent fully-quantum mechanical theoretical results\cite{Klos2023} and, in the case of $^{87}$Rb, the results from experiments utilizing the theory of universality of quantum diffractive collisions.\cite{Booth2019, Shen2020, Shen2021}
For the former, we find excellent agreement; for the latter, we find more nuanced agreement.

The rate coefficient $L(T,W)$ depends on both $T$ and $W$.
The $W$ dependence arises from small angle, glancing collisions that fail to impart enough momentum to eject a cold atom from the trap.  
For small losses due to glancing collisions, we expand
\begin{equation}
  L(T,W) =  K(T) - a_{\rm gl}(T) W + b_{\rm gl}(T)W^2\,,
  \label{eq:loss_rate} 
\end{equation}
where $K(T)$ is the total rate coefficient, $a_{\rm gl}(T)$ and $b_{\rm gl}(T)$ are the first-, and the second-order glancing rate coefficients, respectively.
For convenience, we further define $K(T)= \pazocal{K}_0 + \pazocal{K}_1(T-300\mbox{ K})$, $a_{\rm gl}(T)=\pazocal{A}_0 + \pazocal{A}_1 (T-300\mbox{ K})$, $b_{\rm gl}(T)=\pazocal{B}_0 + \pazocal{B}_1 (T-300\mbox{ K})$ as, in practice, most vacuum chambers operate near ambient temperature.
For the present work, a second-order expansion in $W$ is of sufficient accuracy.

Quantum-mechanical scattering calculations of $K(T)$ and $a_{\rm gl}(T)$,  including an analysis of their theoretical uncertainties, have been conducted for a few systems.
The first to be characterized was $^{6,7}$Li+H$_2$,\cite{Makrides2019, Makrides2022Eb} followed by $^{6,7}$Li+$^4$He.\cite{Makrides2020, Makrides2022Ea}
Recently, Ref.~\onlinecite{Klos2023} presented comprehensive calculations of $^7$Li and $^{87}$Rb colliding with H$_2$, $^{14}$N$_2$, and all the noble gases and provides tables for the coefficients $\pazocal{K}_i$, $\pazocal{A}_i$ and $\pazocal{B}_i$ with $i=0$ or 1.

\begin{figure}
    \centering
    \includegraphics{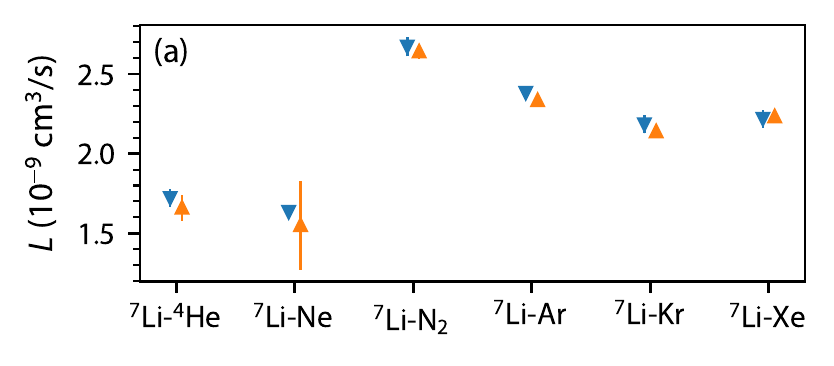}
    \includegraphics{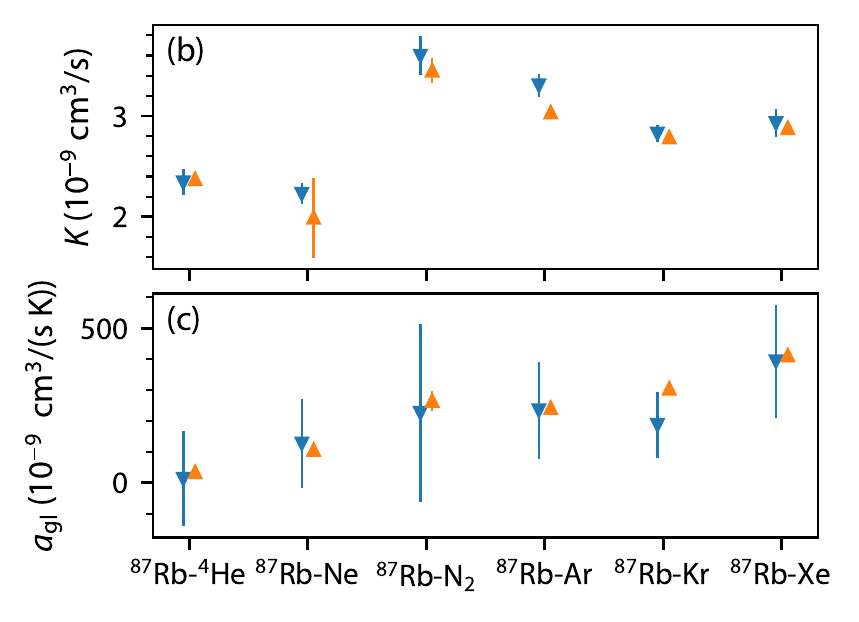}
    \caption{Experimental (blue downward triangles) and theoretical (orange upward triangles) values of $L(W,T)$ in panel (a) for natural abundance noble gas species and nitrogen N$_2$ colliding with $^7$Li  and $K(T)$  and $a_{\rm gl}(T)$ for these same species colliding with $^{87}$Rb in panels (b) and (c), respectively. Error bars are two-standard-deviation combined statistical and systematic ($k=2$) uncertainties.
    The values for $W$ and $T$ can be found in the main text.
    }
    \label{fig:main_results}
\end{figure}

Our principal results for natural abundance noble gas species and nitrogen N$_2$ are given in Fig.~\ref{fig:main_results}.
The figure compares experimentally determined values of $L(W,T)$ for $^7$Li and $K(T)$ and $a_{\rm gl}(T)$ for $^{87}$Rb  with the corresponding theoretical values from Ref.~\onlinecite{Klos2023}.
For $^7$Li data, $W/k=0.95(14)$~mK and $T=300.2(2.9)$~K.
For $^{87}$Rb data, $T=295.2(3)$ K and $W$ ranges between between $k\times0.3986(4)$~mK and $k\times 1.594(1)$~mK in order to extract $K(T)$, $a_{\rm gl}(T)$, and $b_{\rm gl}(T)$.
Most values of $b_{\rm gl}(T)$ are consistent with zero at the two-standard-deviation ($k=2$) level and are thus omitted from Fig.~\ref{fig:main_results}.
The experimental and theoretical values for $L(W,T)$ and $K(T)$ are consistent at the two-standard deviation combined statistical and systematic ($k=2$) uncertainty level, except for $^{7}$Li-Ar and $^{87}$Rb-Ar.

The agreement observed in Fig.~\ref{fig:main_results} has a second or different but equally valid interpretation.
Specifically, the pressure {\it measured} by a cold atom pressure sensor, when using the values of $L$ from Ref.~\onlinecite{Klos2023}, agrees with the pressure {\it set} by a classical dynamic expansion system.
When used to measure pressure in this way, the cold atom pressure sensor is traceable only to the SI second and kelvin, making it a primary standard.
We thus refer to our two sensors as cold atom vacuum standards (CAVSs).
Agreement between a CAVSs and the DE system in our direct comparison validates a CAVS as a standard of vacuum pressure.

Given that a CAVS can easily measure loss rates between $0.01$~s$^{-1}$ and $10$~s$^{-1}$ and typical values of $K(T)$ are on the order of $10^{-9}$~cm$^3$/s at $T=300$ K, we therefore estimate that a CAVS's range of operation spans background-gas number densities (pressures) from of the order of $10^7$~cm$^{-3}$ ($4\times 10^{-8}$~Pa) to $\sim 10^{10}$~cm$^{-3}$ ($4\times 10^{-5}$~Pa).
Indeed, similar devices have been operated up to $6\times 10^{-5}$~Pa.\cite{Shen2020}
These pressures correspond to most of the ultra-high vacuum and part of the high-vacuum regimes.

A significant difference between a CAVS based on $^7$Li and $^{87}$Rb sensor atoms is the value of $a_{\rm gl}(T)$.
$^{87}$Rb with its larger mass has typical values $a_{\rm gl}(T)k\sim 10^{-7}$~cm$^3$/(s~K), while $^7$Li has typical values $a_{\rm gl}(T)k\sim 10^{-8}$~cm$^3$/(s~K).\cite{Klos2023}
In a trap with depth $W \sim k\times 1$~mK, roughly one of every ten collisions between the background gas and a $^{87}$Rb sensor atom is a ``glancing'' collision.
As shown in Fig.~\ref{fig:main_results}, measured values of $a_{\rm gl}(T)$ are consistent at the two-standard deviation combined statistical and systematic ($k=2$) uncertainty level, except for $^{87}$Rb-Kr and $^{87}$Rb-Xe.
For $^7$Li confined in a trap with the same depth, the fractional rate of glancing collisions is an order of magnitude smaller.
Given current fractional measurement uncertainties of order of 1~\%, glancing collisions are thus not detectable for $^7$Li.

The remainder of the paper is divided as follows.
Section~\ref{sec:apparatus} describes the salient features of our two types of apparatuses.
In Sec.~\ref{sec:loss_curves} we analyze our observed sensor atom loss curves as a function of background gas pressure or, equivalently, number density produced by the dynamic expansion system.
Section~\ref{sec:analysis} presents our measured total and glancing rate coefficients along with a description of uncertainty budgets.
We conclude in Sec.~\ref{sec:conclusion}.
Appendices~\ref{sec:apparatus:dynamic_expansion} and~\ref{sec:apparatus:imaging} provide additional details on the dynamic expansion standard and sensor atom imaging, respectively.

\section{Apparatus}
\label{sec:apparatus}
Our apparatuses \cite{Barker2022, Siegel2021, Ehinger2022, Eckel2022} have been described elsewhere.
Briefly, a laboratory-scale cold-atom vacuum standard (l-CAVS),\cite{Barker2022} operating with $^{87}$Rb as its sensor atom, and a portable cold-atom vacuum standard (p-CAVS),\cite{Ehinger2022} operating with $^7$Li as its sensor atom, are attached to a dynamic expansion standard.
The dynamic expansion standard sets a known partial pressure of a gas of interest between $2\times 10^{-8}$~Pa and $2 \times 10^{-6}$~Pa.
In this standard, a known number flow of gas $\dot{N}$, with dimension atoms or molecules per unit time, is injected into a chamber.
This first chamber, to which the CAVSs are attached, connects to a second chamber via a small orifice with a well-characterized flow conductance.
(See Fig.~4 of Ref.~\onlinecite{Barker2022}.)
As shown in Appendix~\ref{sec:apparatus:dynamic_expansion}, the additional number density of atoms or molecules with mass $m$ and at temperature $T$ at the location of the CAVS is
\begin{equation}
    n = \frac{\dot{N}}{\alpha_{\rm MC} A}\sqrt{\frac{2\pi m}{kT}}\frac{r_{\rm p}}{r_{\rm p}-1}\,,\label{eq:n}
\end{equation}
where $\alpha_{\rm MC}$ is the probability of transmission of an atom or molecule through the orifice, $A$ is the opening area of the orifice, and $r_p$ is the measured ratio of pressure in the first chamber to the pressure in the second chamber.
Here, the total gas number density $n_{\rm total} = n + n_{\rm base}$, where $n_{\rm base}$ is the gas number density at base pressure.
For the remainder of this paper, we shall simply call $n$ the number density.

While a known partial number density is generated, either the l-CAVS or the p-CAVS measures the loss rate $\Gamma$ of sensor atoms held in a  quadrupole magnetic trap.
Simultaneous operation of both CAVSs was not possible because operation of the l-CAVS interferes with the stability of the p-CAVS.
Preparation of the sensor atom cloud  in either CAVS involves several steps (see Refs.~\onlinecite{Barker2022, Ehinger2022}).
First, a magneto-optical trap (MOT) is loaded with atoms.
Complementary metal-oxide semiconductor (CMOS) cameras record fluorescence images of the MOT during the loading process and we determine the final number of atoms in the MOT, $N_0$, using these images.
For both the l- and p-CAVSs, $N_0$ is of the order of $10^6$.

Next, the atoms are transferred into the quadrupole magnetic trap. 
For both the l- and p-CAVSs, the transfer process involves optical pumping to the $F=1$ hyperfine ground state and, for the l-CAVS, subsequent heating and removal of any remaining $F=2$ hyperfine states. See Ref.~\onlinecite{Ehinger2022,Barker2022a} for details.
All trapped atoms are then in the  $F=1$, $m_F=-1$ hyperfine state.

Radio frequency (RF) radiation with a frequency $\nu_{\rm RF}$ between 5~MHz and 40~MHz induces spatially localized transitions between magnetic Zeeman states of the sensor atom and sets the energy  depth of the magnetic trap to $W=h\nu_{\rm RF}(1-M g/\mu_{\rm B} g_F B')$, where $M$ is the mass of a sensor atom, $g$ is the local gravitational acceleration, $g_F$ is the Land\'{e} g-factor, and $\mu_{\rm B}$ is the Bohr magneton.
In practice, after loading the l-CAVS magnetic trap, this so-called RF knife is applied with an initial frequency of 40~MHz.
The RF frequency is then linearly decreased to $\nu_{\rm RF}=5$~MHz in 1~s.
The end of this RF frequency ramp corresponds to $t=0$ for the l-CAVS loss rate measurement.
At $t=0$, the remaining $10^5$ $^{87}$Rb atoms have a temperature between 50~$\mu$K and 200~$\mu$K.
The former estimate comes from fitting an {\it in situ} image of the atoms in the magnetic trap to the expected distribution for a thermal cloud; the latter comes from time-of-flight expansion of similarly-prepared clouds with 10 times the atom number to achieve good signal-to-noise.
For $t>0$, the RF frequency is changed to a final, constant $\nu_{\rm RF}$ between 10~MHz and 40~MHz and is applied for the remainder of the time the atoms are in the magnetic trap.
This controllably sets the trap depth to values between $k\times0.3986(4)$~mK and $k\times 1.594(1)$~mK.
We have verified the effectiveness with which our RF knife removes atoms with $E>h\nu_{\rm RF}$ by extending the RF knife ramp down to $\nu_{\rm RF}=100$~kHz, which removes all the atoms.

For the p-CAVS, approximately $1\times 10^5$ $^7$Li atoms are transferred from a grating MOT\cite{Barker2019} into a magnetic quadrupole trap with axial magnetic field gradient $B' = 4.59(17)$~mT/cm. No RF knife is used in the p-CAVS, instead the trap depth is set by the distance between the center of the trap and the nearest in-vacuum surface, the magneto-optical trap's diffraction grating.\cite{Eckel2018}
We calculate a trap depth of $W/k = 0.95(14)$~mK, where the uncertainty comes from the uncertainty in the distance.
The temperature of the magnetically-trapped $^7$Li cloud could not be measured.
It can be as high as $0.75$~mK based on  temperatures observed in other Li grating MOTs.\cite{Barker2019, Barker2022a}
Loading  atoms into the magnetic trap  marks $t=0$ for the p-CAVS loss rate measurement.

For both l- and p-CAVSs, sensor atoms are held in the magnetic trap for a variable amount of time $t>0$, after which the atoms are recaptured into a MOT.
Fluorescence from the MOT is imaged onto CMOS cameras to determine sensor atom number $N_{\rm S}(t)$ as function of time.
The atom-number measurement is destructive, so the atom cloud preparation described above is repeated for each $t$.  
For the l-CAVS, we also repeat the cloud preparation process for each trap depth $W$.
In practice, we measure the ratio $\eta_{\rm S}(t) = N_{\rm S}(t)/N_0$, which reduces our statistical noise by eliminating fluctuations in the atom number loaded into the MOT $N_0$ from one cloud preparation to the next.
Once a full decay curve is measured, taking between 0.25~h and 3~h, the background gas density $n$ is changed and another decay curve is taken.
We do not require an absolute measurement of sensor atom number, so properties of our imaging system, such as the quantum efficiencies of the cameras, do not contribute to our uncertainty budgets, provided such properties do not vary with time.
Details about our imaging system, including its stability and nonlinearity can be found in Appendix~\ref{sec:apparatus:imaging}.
The instability and lack of linearity add a small systematic uncertainty in our final uncertainty budget for the rate coefficients.

\section{Measured Loss Curves}
\label{sec:loss_curves}

\begin{figure}
    \centering
    \includegraphics{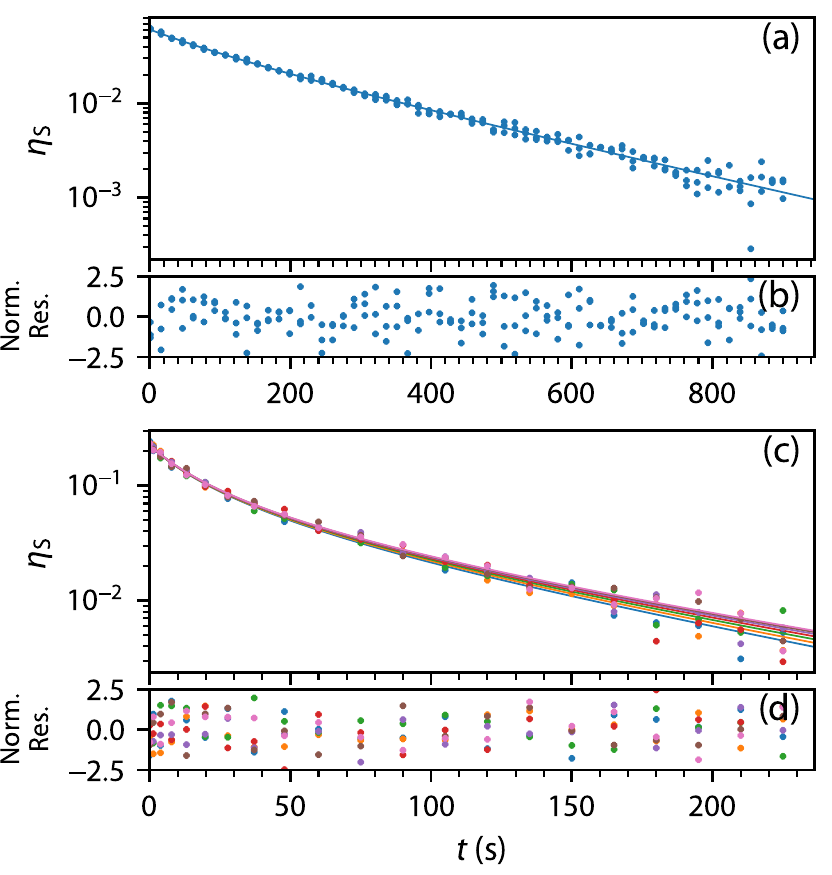}
    \caption{Measured decay curves showing $\eta_{\rm S}(t)$ as functions of hold time $t$ at the lowest achievable pressure in the dynamic expansion system for the p-CAVS with $^7$Li atoms at $T=301.7(3.2)$~K (panels a-b) and the l-CAVS with $^{87}$Rb atoms at $T=295.8(3)$~K (panels c-d). Panels (a) and (c) show the data (colored points) and fits to solutions of Eq.~(\ref{eq:loss}) (solid curves). Panels (b) and (d) show the residuals of the fit normalized to our noise function. For $^{87}$Rb, the colors of the points and curves encode trap depth $W$ with $W/k= 0.3985(4)$~mK (blue), $0.5978(6)$~mK (orange), $0.7970(8)$~mK (green), $0.996(1)$~mK (red), $1.195(1)$~mK (violet), $1.394(1)$~mK (brown), and $1.594(2)$~mK (pink).
    }
    \label{fig:background_decays}
\end{figure}

Before we add background gas to the dynamic expansion system, we record the number of sensor atoms as function of time in the quadrupole magnetic traps of the l-CAVS and p-CAVS at the lowest reachable, or base, pressure (\textit{i.e.} at $n=0$ as defined by Eq.~\ref{eq:n}). 
These decays curves for $\eta_{\rm S}(t) = N_{\rm S}(t)/N_0$ as functions of time $t$ are shown in Fig.~\ref{fig:background_decays}.
For these traces, $T = 295.8(3)$~K for the l-CAVS, and $T = 301.7(3.3)$~K for the p-CAVS.
At base pressure, the decay curves from  both the p-CAVS and l-CAVS are non-exponential.
This non-exponential decay of $N_{\rm S}$ is well described by the solution to the differential equation
\begin{equation}
    \label{eq:loss}
    \frac{{\rm d}\eta_{\rm S}}{{\rm d}t} = -\Gamma \eta_{\rm S} - \beta \eta_{\rm S}^2\,,
\end{equation}
where $\Gamma$ is the trap loss rate and $\beta$ is a two-body loss rate.
We have taken $^{87}$Rb data at several trap depths $W$, so we further parameterize 
\begin{equation}
    \Gamma(W) = \Gamma_0 - \Gamma_1 W + \Gamma_2W^2{\rm \ \ and\ \ } \beta(W)=\beta_0 + \beta_1 W\,.
    \label{eq:Rbparam}
\end{equation}

We find that satisfactory fits to the decays curves can be found by adjusting the initial $\eta_{\rm S}(t=0)$, the parameters in $\Gamma(W)$ and $\beta(W)$ as well as  $\sigma_\eta$ and $\sigma_0$ in noise function $u(\eta_{\rm S}) = \sqrt{(\sigma_\eta \eta_{\rm S})^2 + \sigma_0^2}$.
The noise function is a model for the uncertainty in the sensor atom number and is an implicit function of time $t$.
The first component, proportional to $\eta_{\rm S}$, is related to random fluctuations in the initial sensor atom number in the magnetic trap and the fluctuating detuning of the MOT laser beams.
The second component $\sigma_0$ reflects the minimum number of sensor atoms that is detectable by our imaging system.
The parameters $\sigma_\eta$ and $\sigma_0$ are different for $^7$Li and $^{87}$Rb but should be independent of background species, $n$, and $W$.

For $^7$Li in the p-CAVS with its fixed $W$, we fit all values for $\eta_{\rm S}(t)$ to Eq.~(\ref{eq:loss}) and, in this manner, determine $\Gamma$ and $\beta$ and their covariances.
For $^{87}$Rb in the l-CAVS with its variable $W$, we simultaneously fit  the time traces $\eta_{\rm S}(t)$ at all $W$ to the combination of Eqs.~(\ref{eq:loss}) and (\ref{eq:Rbparam}).
This procedure gives us reliable values for the two parameters in the noise function, as a single time trace at a single $W$ does not contain enough data.
This simultaneous fit determines $\Gamma_0$, $\Gamma_1$, $\beta_0$, and $\beta_1$ and their covariances.

Figure~\ref{fig:background_decays} also shows the quality of our fits. 
The residuals normalized by the noise function do not have recognizable patterns.
A cumulative distribution function (CDF) constructed from the residuals is well described by the CDF for a Gaussian distribution.
For our p-CAVS  with $^7$Li atoms $\sigma_\eta\lesssim 0.03$, while for our l-CAVS with $^{87}$Rb atoms,  $\sigma_\eta\lesssim 0.08$.
The minimum detectable atom number $\sigma_0$ is about 500 for the p-CAVS and is $300$ for the l-CAVS.

The best fit values of $\Gamma(W_{\rm p-CAVS})$ for $^7$Li and $\Gamma_0$ for $^{87}$Rb are 0.00388(6) s$^{-1}$ and 0.0119(8) s$^{-1}$, respectively.
Here, $W_{\rm p-CAVS}$ is the fixed trap depth of the p-CAVS.
Assuming that H$_2$ is our dominant background gas and using the theoretical values of rate coefficients $K_{{\rm Li-H}_2}=3.18(6)\times 10^{-9}$ cm$^3$/s at $T = 301.7(3.3)$~K and $K_{{\rm Rb-H}_2}= 3.9(1)\times 10^{-9}$ cm$^3$/s at $T = 295.8(3)$~K, we find pressures of 5.19(3)~nPa and 14.2(1.4)~nPa, according to $^7$Li p-CAVS and $^{87}$Rb l-CAVS, respectively.
Here, the uncertainty is dominated by the uncertainty in the theoretical rate coefficients.
The factor of nearly three difference in the base pressure readings may be due to a variety of factors, including pressure gradients (see Appendix~\ref{sec:apparatus:dynamic_expansion}), the difference in Majorana loss of the two species, and the inability to accurately separate $\Gamma$ from  two- or even three-body losses in the fits.
We note that a previous experiment with two p-CAVSs closely connected to each other on a different vacuum chamber than used here measured the same, higher pressure (42.2(1.0)~nPa) within their respective uncertainties.\cite{Ehinger2022}

For $^{87}$Rb, we find $\Gamma_1={\rm d}\Gamma/{\rm d}W = 1.56(81)$~s$^{-1}$/K. 
This value is consistent with zero at two standard deviations ($k=2$).
The ratio of $\Gamma_1/\Gamma_0 = 142(85)$~K$^{-1}$ is likewise consistent with the theoretical prediction of 36.7(1.8)~K$^{-1}$ for $^{87}$Rb+H$_2$ and a recent measurement\cite{Shen2022} of $43(5)$~K$^{-1}$.

We convert the fitted values of $\beta (W_{\rm p-CAVS})$ for $^7$Li and $\beta_0$ for $^{87}$Rb from the data in Fig.~\ref{fig:background_decays} to rate coefficients $K_2$ defined through the differential equation $\dot{n}_{\rm S} = -\Gamma n_{\rm S} - K_2 n_{\rm S}^2$ for the sensor atom number density $n_{\rm S}(t)$.\cite{Yan2011, Barker2022}
The fitted $\beta_1$ is consistent with zero.
For $^{87}$Rb, the derived $K_2\approx 2\times10^{-10}$~cm$^3$/s is remarkably close to the known elastic scattering rate coefficient of $1.2\times 10^{-10}$~cm$^3$/s among $^{87}$Rb atoms using the {\it in situ} rubidium temperature estimate of 50~$\mu$K.\cite{Barker2022}
Elastic collisions only change the momenta of the atoms and thus should not lead to sensor atom loss when the sensor atom temperature is much less than the trap depth $W$ as is the case in our $^{87}$Rb experiments.
We observe no difference in the two-body loss rate when we reduce the efficiency of the RF knife by halving the amplitude of the RF radiation, further indicating that our RF knife is efficient at removing highly energetic atoms, which, if left behind, could increase the observed two-body loss rate.
For $^7$Li, the derived $K_2$ is inconsistent and much larger than the known elastic scattering rate coefficient at a lithium temperature of roughly 750~$\mu$K. 
The origin of the non-zero values for $\beta$ in both CAVSs remains a mystery.

\begin{figure}
    \centering
    \includegraphics{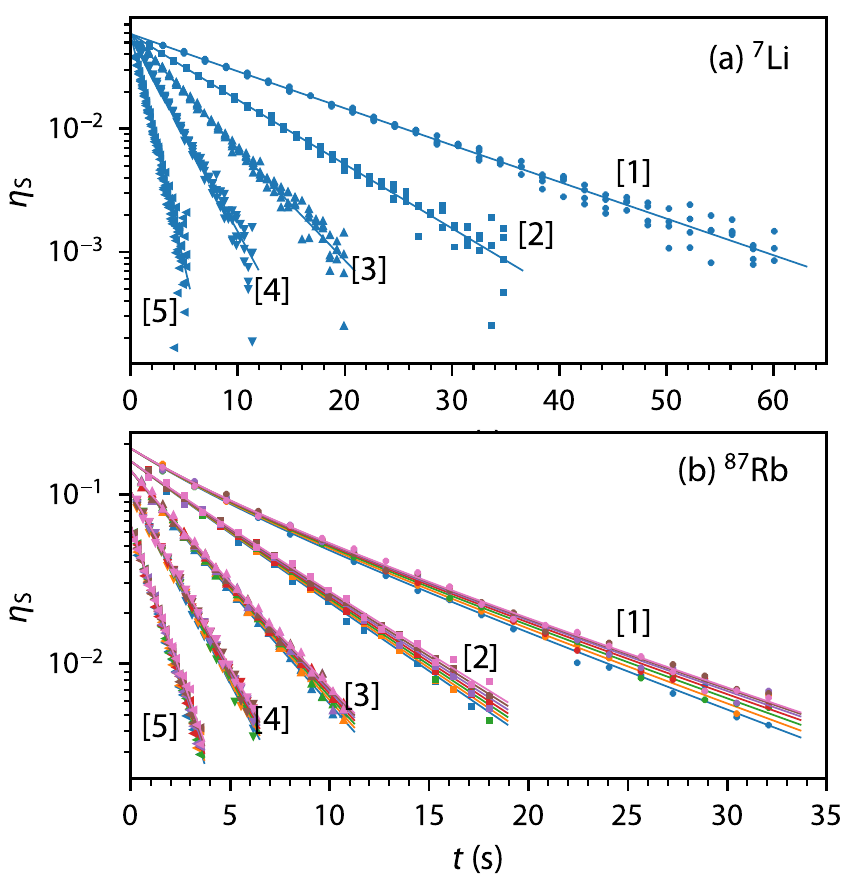}
    \caption{ Measured decay curves of $\eta_{\rm S}(t)$ as functions of time $t$ for several natural-abundance argon background gas number densities $n$ for $^7$Li  (panel a) and $^{87}$Rb (panel b).  For the $^7$Li data, the Ar mean gas temperature is $T=300.7(3.0)$~K and the densities are $2.63(3)\times10^{7}$~cm$^{-3}$ (labeled [1]), $4.75(6)\times10^{7}$~cm$^{-3}$ ([2]), $8.61(9)\times10^{7}$~cm$^{-3}$ ([3]),  $1.50(2)\times10^{8}$~cm$^{-3}$ ([4]), and $3.54(4)\times10^{8}$~cm$^{-3}$ ([5]).  For the $^{87}$Rb data, the Ar mean gas temperature is $T=295.1(3)$~K and the densities are $2.66(2)\times10^{7}$~cm$^{-3}$ ([1]), $4.79(2)\times10^{7}$~cm$^{-3}$ ([2]), $8.73(3)\times10^{7}$~cm$^{-3}$ ([3]),  $1.524(5)\times10^{8}$~cm$^{-3}$ ([4]), and $2.682(9)\times10^{8}$~cm$^{-3}$ ([5]).
    For $^{87}$Rb, the  points and curves with different colors correspond to data taken at different trap depths $W$. The color coding and values for $W$ are  as in Fig.~\ref{fig:background_decays}.
    }
    \label{fig:example_data}
\end{figure}

We are now ready to study the readings of the CAVSs when a known background number density $n$ of a  gas species $X$ is set by the combined flowmeter and dynamic expansion system.
A sampling of the available data for $^7$Li with natural abundance Ar gas, taken with the p-CAVS, and for $^{87}$Rb also with natural abundance Ar gas, taken with the l-CAVS, are shown in Fig.~\ref{fig:example_data}(a) and (b), respectively.
The figure shows $\eta_{\rm S}(t)$ as functions of time for several values of $n$ between $0.2\times 10^8$ cm$^{-3}$ and $4\times 10^8$ cm$^{-3}$.
For $^{87}$Rb, Fig.~\ref{fig:example_data}(b) also shows time traces for several trap depths $W$.
We observe that for roughly the same Ar gas density, the observed lifetimes for $^7$Li are about 60~\% longer than those of $^{87}$Rb, consistent with the observation that the rate coefficients $K$ for $^7$Li are about 60~\% smaller than those of $^{87}$Rb.
We have similar quality data for the other noble gases as well as for N$_2$.
In all cases, we use gases containing  a natural abundance distribution of the stable isotopes.

For $^7$Li, we fit all values for $\eta_{\rm S}(t)$ taken at number density $n$ to Eq.~(\ref{eq:loss}), even though the non-exponential decay is not always apparent.
In this manner, we determine $\Gamma$ and $\beta$ and their covariances for each $n$ and each background gas species $X$.
For $^{87}$Rb, we fit all values of $\eta_{\rm S}(t)$ at all values of $W$ at a single background-gas number density $n$ to the combination of Eqs.~(\ref{eq:Rbparam}) and (\ref{eq:loss}), even though, again, the non-exponential decay is not always apparent.
This simultaneous fit determines $\Gamma_0$, $\Gamma_1$, $\beta_0$, and $\beta_1$ and their covariances for each $n$ and each background gas species $X$.
We find that within their uncertainties the fitted values for $\sigma_\eta$ and $\sigma_0$ are consistent for all $n$ and all background species,
as expected, for both $^7$Li and $^{87}$Rb.

\section{Analysis \& Discussion}

\label{sec:analysis}
\begin{figure}
    \centering
    \includegraphics{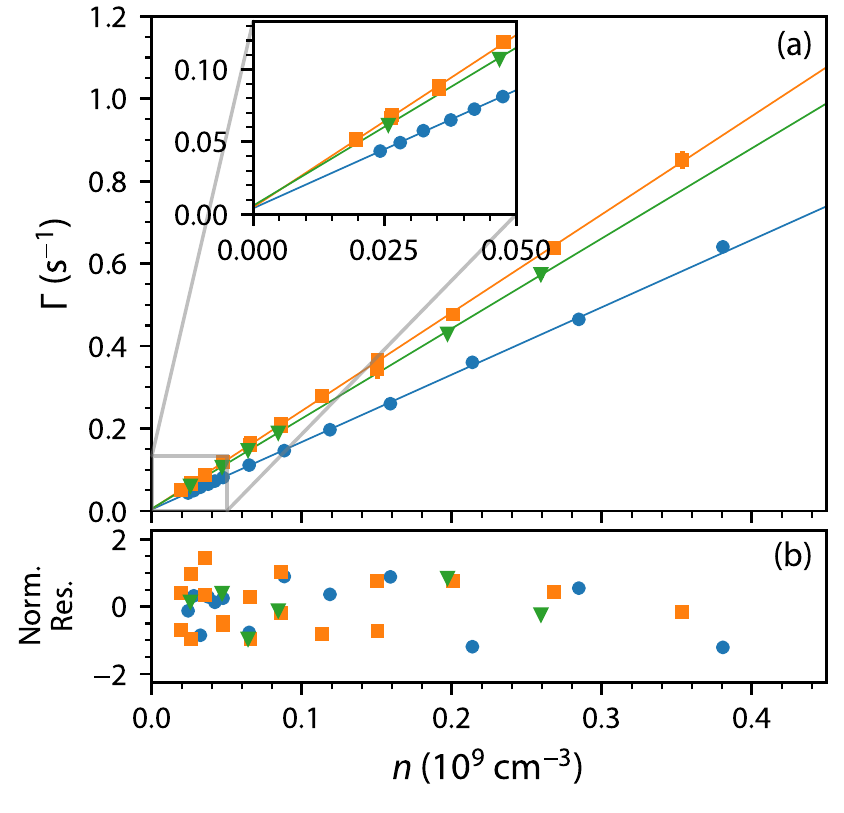}
    \caption{(a) Sensor atom decay rates $\Gamma$ as functions of background gas number density $n$ with linear fits for $^7$Li colliding with Ne (blue circles), Ar (orange squares) and Kr (green trangles) at $T=299.8(2.8)$~K.
    Error bars on $\Gamma$ and $n$ are smaller than the points.
    As shown in the inset, the typical fitted value of the atom loss rate at $n=0$ (i.e., base pressure) is $\Gamma_{\rm base} \approx 0.005$~s$^{-1}$.
    Panel (b) shows the normalized residuals of the linear fit.
    }
   \label{fig:example_analysis}
\end{figure}

The  values for rate $\Gamma$ extracted from fitting $^{7}$Li-atom decay curves for approximately seven background gas number densities $n$ for each background species determine the corresponding rate coefficient $L$. 
These data are uncorrelated.
Figure \ref{fig:example_analysis} shows $\Gamma$ as a function $n$ for natural-abundance background gas species Ne, Ar, and Kr.
The smallest $n$ shown in the figure correspond  to pressures that are still well above our base pressure.
We observe that the $n$-dependence of $\Gamma$  must be described by
\begin{equation}
\label{eq:lin_L}
\Gamma = L n + \Gamma_{\rm base},
\end{equation}
with non-negligible offset rate $\Gamma_{\rm base}$ representing sensor atom loss at base pressure.
In this section, we will use $n=0$ to represent the background gas number density at base pressure.
The $y$-uncertainties of the data in Fig.~\ref{fig:example_analysis} are  the statistical uncertainties of the fitted value of $\Gamma$.
The $x$-uncertainties in the data are due to  combined type-A and type-B uncertainties in $n$, described in Appendix~\ref{sec:apparatus:dynamic_expansion}.
Typically, $u(n)/n \ll u(\Gamma)/\Gamma$.
We fit the data in Fig.~\ref{fig:example_data} to Eq.~(\ref{eq:lin_L}), with each point weighted by variance $\sigma^2 =  u^2_{\rm A}(\Gamma) + L^2 u^2_{\rm A}(n)$, where $u_{\rm A}(O)$ is the statistical (type-A) uncertainty in observable $O$.
Type-B uncertainties are propagated separately.
The value of $\Gamma$ at base pressure, $n=0$, is excluded in the fits for three reasons:
(1) the day-to-day fluctuations in the $\Gamma$ measured at $n=0$, using data similar to Fig.~\ref{fig:background_decays}, are much larger than the statistical uncertainty from the fit to Eq.~(\ref{eq:loss});
(2) inclusion of the measured $\Gamma$ at $n=0$ weighted by its uncertainty $u(\Gamma)$ causes correlations in the residuals of the linear fits; and
(3) we lack confidence that the non-linear least squares fitting algorithm employed in Sec.~\ref{sec:loss_curves} is  accurately separating $\Gamma$ and $\beta$, which itself might indicate that term $\beta \eta_{\rm S}^2$ in Eq.~(\ref{eq:loss}) may not be the correct functional form.

Our values of $\chi^2/\nu$, where $\nu$ is the number of degrees of freedom, are 0.41 ($^7$Li+He, $\nu=5$), 0.59 ($^7$Li+Ne, $\nu=11$), and 0.57 ($^7$Li+N$_2$, $\nu=8$), 0.62 ($^7$Li+Ar, $\nu=16$), 0.47 ($^7$Li+Kr, $\nu=4$), and 1.61 ($^7$Li+Xe, $\nu=5$).
In fact, no fits fail the $\chi^2$ test,\cite{Bevington1992} where the probability of a hypothetical repeated realization of the experiment with the same uncertainties yielding a larger $\chi^2/\nu$ is less than 5~\%.
Fitted values of $\Gamma_{\rm base}$ range from 0.0042(6)~s$^{-1}$ to 0.066(4) s$^{-1}$, consistent with the long-term fluctuations observed in repeated measurements of the decay rate at base pressure (such as that shown in Fig.~\ref{fig:background_decays}).

\begin{table}
\begin{tabular}{lclr}
\hline\hline 
 & \multicolumn{1}{c}{Type} & \multicolumn{1}{c}{Source} &  Contribution (\%) \\ 
 \hline
Experi- & B & Temperature of the CAVS, $T$ & 0.51 \\
\ \ mental & B & Flowmeter, $\dot N$ & 0.24 \\ 
 & B & Orifice Area, $A$ & 0.13 \\ 
 & B & Imaging non-linearity \& stability, $\Gamma$ & 0.07 \\ 
 & B & Pressure Ratio, $r_{\rm p}$ & 0.05 \\ 
 & B & Orifice Transmission Prob., $\alpha_{\rm MC}$ & 0.02 \\ 
\hline 
 & B & Subtotal & 0.58 \\ 
 & A & Subtotal & 0.57 \\ 
\hline 
 & & Total & 0.82 \\ 
\hline\hline 
Theory & B & Temperature of the CAVS, $T$ & 0.28 \\
 & B & Theory Coefficients, $K$ & 0.25 \\ 
 & B & Trap Depth, $W$ & 0.07 \\ 
 & B & Isotopic composition & $<0.01$ \\
 & B & Temperature of cold atoms, $T_c$ & $<0.01$ \\
\hline 
 & & Total & 0.41 \\ 
\hline\hline 
\end{tabular}
\caption{Experimental and theoretical statistical (type-A) and systematic (type-B)   uncertainty budgets of loss rate coefficient $L$ for the p-CAVS with $^7$Li sensor atoms and a natural abundance Ar background gas.  The experimental and theoretical contributions add in quadrature to the {\it relative} uncertainty $u(L)/L$ of  $L$. The experimental and theoretical temperature contributions are correlated. See text on how this non-zero correlation is treated.}
\label{tab:uncertainty_budget}
\end{table}

We can now discuss the systematic, type-B uncertainties of the data for $^7$Li in Fig.~\ref{fig:example_analysis}.
These are 
(a) the uncertainty in the measured flow, which has a complicated dependence on $\dot{N}$,\cite{Eckel2022}
(b) the uncertainties in the orifice transmission $\alpha_{\rm MC}$ and area $A$,
(c) the uncertainty in the fitted value of $\Gamma$ due to the imaging non-linearities and stability,
(d) the uncertainty in the measurement of $r_{\rm p}$, and 
(e) the uncertainty in the measurement of the background gas temperature $T$.
For pairs of observables $O$ and $P$ with $O,P\in\{\dot{N},  \alpha_{\rm MC}, A, \Gamma, r_{\rm p}, T\}$, we chose the covariance matrix for these type-B uncertainties to be equal to
\[
   {\rm cov}(O,P)=u^2_{\rm B}(O)\delta_{O,P}\,,
\]
where $\delta_{O,P}=1$ for $O=P$ and 0 otherwise.
The type-B standard uncertainty of observable $O$ is
\begin{equation}
    u_{\rm B}(O)=\frac{\sum_i u(O_{i})/\sigma_i^2}{\sum_i 1/\sigma_i^2},
\end{equation}
where index $i$ labels data points $(\Gamma_i,n_i)$ of independently extracted $\Gamma_i$ at number density $n_i$.
Then, $u(O_i)$ is the standard uncertainty of observable $O$ recorded during the taking of data point $i$, and  $\sigma_i^2 =  u^2_{\rm A}(\Gamma_i) + L^2 u^2_{\rm A}(n_i)$ is the type-A variance at data point $i$.

The type-B uncertainty of $L$ with measurement equation
\begin{equation}
    L = \frac{\Gamma}{n} = \frac{\Gamma \alpha_{\rm MC} A}{\dot{N}}\sqrt{\frac{kT}{2\pi m}}\frac{r_{\rm p}-1}{r_{\rm p}}
    \label{eq:typeB}
\end{equation}
from Eq.~(\ref{eq:n}) and $\Gamma = L n$ then follows from standard error propagation using ${\rm cov}(O,P)$.

Table~\ref{tab:uncertainty_budget} shows the complete uncertainty budget for the experimental value of $L$ for $^7$Li+Ar.
Its statistical uncertainty follows from the linear least squares fit for $L$ of the data in Fig.~\ref{fig:example_analysis}.
We observe that the statistical and systematic uncertainties of the experimental $L$ are approximately equal.
The experimental uncertainty budgets for $L$ of $^7$Li with other natural-abundance background species are similar.

Atom loss decay curves for $^{87}$Rb sensor atoms described in the previous section
have resulted in values for $\Gamma_0$ and $\Gamma_1$ at approximately ten background number densities $n$ for each background species. The values for $\Gamma_0$ and $\Gamma_1$ at the same $n$ and background species are correlated.
The approximately ten values of $\Gamma_0$ are then fit
to  $\Gamma_0 = K n + \Gamma_{0,{\rm base}}$ and we find values, uncertainties, and covariances for $K$ and $\Gamma_{0,{\rm base}}$.
Finally, we fit all values for $\Gamma_1$ to  $\Gamma_1 = a_{\rm gl} n + \Gamma_{1,{\rm base}}$ and all values for $\Gamma_2$ to  $\Gamma_2 = b_{\rm gl} n + \Gamma_{2,{\rm base}}$
and obtain $a_{\rm gl}$, $\Gamma_{1,{\rm base}}$, $b_{\rm gl}$, $\Gamma_{2,{\rm base}}$, respectively.
As for $^7$Li, we do not include data taken at base pressure in the fits to determine these four parameters.
We find that the values of $\chi^2/\nu$ for the linear least squares fits to extract $K$ are 1.32 ($^{87}$Rb+He, $\nu=8$), 1.20 ($^{87}$Rb+Ne, $\nu=7$), and 1.07 ($^{87}$Rb+N$_2$, $\nu=8$), 0.33 ($^{87}$Rb+Ar, $\nu=7$), 0.42 ($^{87}$Rb+Kr, $\nu=6$), and 1.34 ($^{87}$Rb+Xe, $\nu=8$).
Again, no fits fail the $\chi^2$ test. 

Values of $\Gamma_{0,{\rm base}}$ for $^{87}$Rb range from 0.017(2)~s$^{-1}$ to 0.034(4) s$^{-1}$, much larger than rate 0.0119(8)~s$^{-1}$ determined from the fit to data shown in  Fig.~\ref{fig:background_decays}(b).
This larger $\Gamma_{0,\rm base}$ suggests that we can not sufficiently separate the effects from $\beta \eta_{\rm S}^2(t)$ and $\Gamma \eta_{\rm S}(t)$ in decay curves.
We have performed analyses of the experimental systematic uncertainties in the same manner as described for the $^7$Li data.
Our systematic relative uncertainties are approximately the same as those for the $^7$Li experiments.
The relative statistical uncertainties of $K$, however, are larger by a factor between 2 and 4 compared to those for $L$ of $^7$Li. 

\begin{table}
\begin{tabular}{lD{.}{.}{1.4}D{.}{.}{1.4}D{.}{.}{1.2}}
\hline\hline
System  & \multicolumn{1}{c}{$L$ (thr)} & \multicolumn{1}{c}{$L$ (exp)} & \multicolumn{1}{c}{$E(L)$} \\
 & \multicolumn{1}{c}{($10^{-9}$cm$^3$/s)} &  \multicolumn{1}{c}{($10^{-9}$cm$^3$/s)} \\
\hline
 $^7$Li-$^4$He & 1.66(4) & 1.72(3) & 0.64 \\
 $^7$Li-Ne & 1.6(1) & 1.63(2) & 0.30 \\
 $^7$Li-N$_2$ & 2.64(2) & 2.67(3) & 0.45 \\
 $^7$Li-Ar & 2.34(1) & 2.38(2) & 1.15 \\
 $^7$Li-Kr & 2.140(7) & 2.18(3) & 0.84 \\
 $^7$Li-Xe & 2.23(2) & 2.22(3) & -0.23 \\
\hline
\end{tabular}
\caption{Theoretical (thr) and experimentally (exp) determined values of the total loss rate $L$ for various natural abundance gases colliding with ultracold $^{7}$Li.
The degree of equivalence is $E_n(L) = (L_{\rm exp} - L_{\rm thr})/[2 u(L_{\rm exp}-L_{\rm thr})]$. 
All uncertainties are one-standard deviation $k=1$ uncertainties.}
\label{tab:Li_result}
\end{table}

We now determine the systematic uncertainty budgets of the theoretical expectations for $L$, $K$, $a_{\rm gl}$ and $b_{\rm gl}$ given the experimental conditions and data in Ref.~\onlinecite{Klos2023}.
For $^7$Li-$X$ systems, we evaluate Eq.~(\ref{eq:loss_rate}) at the experimental values for temperature $T$ and trap depth $W$ and account for their uncertainties.
For $^{87}$Rb, we only need to evaluate $K(T)$ at the experimental temperature and account for its uncertainty.
Note that the theoretical uncertainty $u_{\rm B}(T)$ is the same as that  used to determine the uncertainties of conductance $C_0$ and thermal transpiration effects in Eq.~(\ref{eq:typeB}).
Hence, the theoretical (thr) and experimental (exp) rate coefficients are correlated with covariance
\begin{equation}
    {\rm cov}(L_{\rm thr},L_{\rm exp})  = \frac{\partial L_{\rm thr}}{\partial T} u^2_{\rm B}(T)\frac{\partial L_{\rm exp}}{\partial T}= \frac{1}{2}  \pazocal{K}_1    \frac{L_{\rm exp}}{T}       u^2_{\rm B}(T) \,.
    \label{eq:covariance}
\end{equation}
The covariance ${\rm cov}(K_{\rm thr},K_{\rm exp})={\rm cov}(L_{\rm thr},L_{\rm exp})$, and ${\rm cov}(a_{\rm gl, thr},a_{\rm gl, exp})$ is the same as Eq.~(\ref{eq:covariance}) with $\pazocal{K}_1$ replaced by $\pazocal{A}_1$.

In addition, we must adjust for the fact that we experimentally use natural isotope abundance background gases while the data in Ref.~\onlinecite{Klos2023} is computed for a gas containing only the most abundant isotope.
We  scale the theoretical rate coefficients for one isotope to values for other isotopes by using the semiclassical dependence on the mass of the background gas species $m$ and the mass of the sensor atom $M$.
We then find ``weighted'' rate coefficients  based on the natural abundance of each isotope.
The relevant semi-classical mass dependencies are $K\propto m^{-3/10}$, $a_{\rm gl}\propto m^{-1/10}$, and $b_{\rm gl}\propto m^{1/10}$.
This scaling matters most for neon and xenon, for which the isotope correction $\delta K_{\rm isotope}$ represents a $-0.28$~\% and $+0.14$~\% shift in $K$, respectively.
We take this scaling to be approximate with a 15~\% relative uncertainty that is $u(K)|_{\rm isotope}=0.15\,\delta K_{\rm isotope}$ to be added in quadrature to all other uncertainties in the theoretical rate coefficients.
The relative uncertainty due to isotopic abundance for $b_{\rm gl}$ is negligible, so we omit it from the uncertainty budget.

We also consider the effect of the temperature of the cold atom cloud, $T_c$, on the theoretical prediction.
Reference~\onlinecite{Klos2023} computes its results using a reference value $T_{c0}=100$~$\mu$K; nonzero differences $T_c-100 \mbox{ $\mu$K}$ are accounted for by estimating the change of the effective collision temperature\cite{Makrides2019}
\begin{equation}
    T_{\rm eff} = \frac{M}{m+M}T + \frac{m}{m+M}T_c\,,
\end{equation}
which leads to the modified first-order expansion\cite{Makrides2019,Makrides2020}
\begin{equation}
    K(T) = \pazocal{K}_0 + \pazocal{K}_1(T-300\mbox{ K}) + \pazocal{K}_1 \frac{m}{M}(T_c-100 \mbox{ $\mu$K})\,.
\end{equation}
For both the p- and the l-CAVS, we use $T_c=100$~$\mu$K, and assume symmetric uncertainties for simplicity.
For the p-CAVS, we take $u(T_c)=350$~$\mu$K, which encompasses the $750$~$\mu$K maximum temperature at $k=2$; for the l-CAVS, we take $u(T_c)=50$~$\mu$K, which encompasses the $50$~$\mu$K to $200$~$\mu$K range at $k=2$.
For both the p- and the l-CAVS, the additional relative uncertainty to $L$ is $<0.01$~\%, significantly smaller than many other sources of uncertainty.
We include it in the uncertainty budget for completeness.

Table~\ref{tab:uncertainty_budget} shows the uncertainty budget  in the theoretical value for $L$ for the $^7$Li+Ar system.
The relative uncertainty for the theoretical value is half that of the combined systematic and statistical uncertainties of the experimental value.
Table~\ref{tab:Li_result} shows our final theoretical and experimental values of $L$ for $^7$Li+$X$ systems, along with the degree of equivalence $E(L)$ for $L$ defined by  $E(O) = (O_{\rm thr} - O_{\rm exp})/[2u(O_{\rm thr} - O_{\rm exp})]$, where $u(O_{\rm thr} - O_{\rm exp})=\sqrt{u^2(O_{\rm thr})-2\,{\rm cov}(O_{\rm thr}, O_{\rm exp})+u^2(O_{\rm exp})}$
is the uncertainty of the difference between the correlated theoretical  and experimental values for quantity $O$.
As the temperature dependence of the theory and experiment values are correlated, $E(L)$ is larger  than the uncorrelated combination of the theoretical and experimental uncertainties would suggest.
All values agree at  three standard deviations, $k=3$, all except $^7$Li-Ar agree at $k=2$.

\begin{table*}
\begin{tabular}{lD{.}{.}{1.4}D{.}{.}{1.4}D{.}{.}{3.4}D{.}{.}{1.5}D{.}{.}{1.6}D{.}{.}{3.4}D{.}{.}{1.5}D{.}{.}{1.4}D{.}{.}{3.4}}
\hline\hline
System  & \multicolumn{1}{c}{$K$ (thr)} & \multicolumn{1}{c}{$K$ (exp)} & \multicolumn{1}{c}{$E(K)$} & \multicolumn{1}{c}{$a_{\rm gl}$ (thr)} & \multicolumn{1}{c}{$a_{\rm gl}$ (exp)} & \multicolumn{1}{c}{$E(a_{\rm gl})$} & \multicolumn{1}{c}{$b_{\rm gl}$ (thr)} & \multicolumn{1}{c}{$b_{\rm gl}$ (exp)} & \multicolumn{1}{c}{$E(b_{\rm gl})$} \\
 & \multicolumn{1}{c}{($10^{-9}$cm$^3$/s)} &  \multicolumn{1}{c}{($10^{-9}$cm$^3$/s)} & & \multicolumn{1}{c}{($10^{-7}$cm$^3$/[s K])} & \multicolumn{1}{c}{($10^{-7}$cm$^3$/[s K])} & & \multicolumn{1}{c}{($10^{-5}$cm$^3$/[s K$^2$])} & \multicolumn{1}{c}{($10^{-5}$cm$^3$/[s K$^2$])} & \\
\hline
$^{87}$Rb-$^4$He & 2.37(3) & 2.34(6) & -0.23 & 0.336(5) & 0.12(77) & 0.14 & 0.067(3) & \multicolumn{1}{c}{---} & \multicolumn{1}{c}{---} \\
 $^{87}$Rb-Ne & 2.0(2) & 2.23(5) & 0.58 & 1.06(9) & 1.27(72) & -0.15 & 0.59(3) & 2.6(2.4) & -0.41 \\
 $^{87}$Rb-N$_2$ & 3.45(6) & 3.6(1) & 0.64 & 2.6(2) & 2.3(1.4) & 0.13 & 2.3576(7) & 13.3(8.4) & -0.66 \\
 $^{87}$Rb-Ar & 3.035(7) & 3.30(6) & 2.27 & 2.42(2) & 2.34(79) & 0.05 & 2.19(2) & -1.0(2.9) & 0.54 \\
 $^{87}$Rb-Kr & 2.79(1) & 2.83(4) & 0.47 & 3.04(2) & 1.87(53) & 1.11 & 3.97(3) & 3.8(2.7) & 0.03 \\
 $^{87}$Rb-Xe & 2.88(1) & 2.93(7) & 0.35 & 4.11(5) & 3.93(92) & 0.10 & 7.1(1) & 10.3(3.0) & -0.53 \\
\hline
\end{tabular}
\caption{
Theoretically (thr) and experimentally (exp) determined values of the loss rate coefficient $K$ at zero trap depth, the first-order glancing rate coefficient $a_{\rm gl}$, and the second-order glancing rate coefficient $b_{\rm gl}$ for various natural abundance  gases colliding with ultracold $^{87}$Rb.
Numbers in parentheses are one-standard-deviation, $k=1$ uncertainties.
The degree of equivalence is $E_n(K) = (K_{\rm exp} - K_{\rm thr})/[2 u(K_{\rm exp}-K_{\rm thr})]$ for $K$ and likewise for $a_{\rm gl}$.}
\label{tab:Rb_results}
\end{table*}

Table~\ref{tab:Rb_results} shows the predicted and measured $K$, $a_{\rm gl}$, and $b_{\rm gl}$ for $^{87}$Rb colliding with He, Ne, N$_2$, Ar, Kr and Xe as background species.
We find $k=2$ agreement between the theoretical and experimental $K$ for all collision partners except $^{87}$Rb+Ar.
The theoretical and experimental values of $a_{\rm gl}$ and $b_{\rm gl}$ agree at $k=2$ for all collision partners except $^{87}$Rb+Kr, which agrees at $k=3$.
We constrained $\Gamma_2=0$ in our fits $^{87}$Rb+He because the expected size of $b_{\rm gl}$ is two orders of magnitude lower than the uncertainty on the values for all other background species.
The experimental relative uncertainties for $a_{\rm gl}$ are much larger than the corresponding theoretical uncertainties and experimental uncertainties observed in Ref.~\onlinecite{Booth2019, Shen2020, Shen2021} because the present experiment focused on taking data at many distinct pressures, rather than at many trap depths for each pressure.

We examined several other potential systematic effects.
For the l-CAVS, we studied sensor atom loss rates after changing the laboratory temperature from 22.0(1)~$^\circ$C to 19.0(5)~$^\circ$C, the magnetic field gradient of the quadrupole trap from 18~mT/cm to 9.0~mT/cm and 24~mT/cm, and the applied RF powers from 25~W to 12~W, but saw no statistically significant dependence of $K$ or $a_{\rm gl}$ on these parameters.
We also tested an alternative application of the RF knife, that of Ref.~\onlinecite{Booth2019}.
After loading the magnetic quadrupole trap and waiting for a time $t$, we apply an RF sweep such that the trap depth decreases from $k\times 3.188(3)$~mK to final trap depth $W$ to eject sensor atoms with kinetic energy $E>W$ and, immediately afterward, measure the final atom number $N$.
We observed no change of $K$ or $a_{\rm gl}$ when using this alternative application of the RF knife.
For the p-CAVS, we changed the power dissipated in the source from 2.7~W to 2.0~W and 3.5~W, magnetic field gradient of the quadrupole trap from 4.59~mT/cm to 7.53~mT/cm, and laboratory temperature from 22.0(1) $^\circ$C to 19.0(5)~$^\circ$C  and 25.0(1)~$^\circ$C, but again saw no statistically significant dependence of $L$ on these parameters.

\begin{table}
\begin{tabular}{lD{.}{.}{2.6}D{.}{.}{1.4}D{.}{.}{2.6}D{.}{.}{1.4}}
\hline\hline
System  & \multicolumn{4}{c}{$K$ ($10^{-9}$cm$^3$/s)} \\
 & \multicolumn{1}{c}{UQDC\cite{Shen2021}} & \multicolumn{1}{c}{Ratiometric\cite{Shen2022}} & \multicolumn{1}{c}{Theory\cite{Klos2023}} &\multicolumn{1}{c}{This work}  \\
\hline 
 $^{87}$Rb-H$_2$ &  5.12(15) & 3.8(2) &  3.9(1) &\multicolumn{1}{c}{---} \\
 $^{87}$Rb-$^4$He &  2.41(14) & \multicolumn{1}{c}{---}  & 2.37(3) & 2.34(6) \\
 $^{87}$Rb-Ne &  \multicolumn{1}{c}{---} & \multicolumn{1}{c}{---}  & 2.0(2) & 2.23(5) \\
 $^{87}$Rb-N$_2$ & 3.14(5) & \multicolumn{1}{c}{---}  & 3.45(6) &3.6(1) \\
 $^{87}$Rb-Ar &  2.79(5) & \multicolumn{1}{c}{---}  & 3.035(7) & 3.30(6) \\
 $^{87}$Rb-CO$_2$ & 2.84(6) & \multicolumn{1}{c}{---}  & \multicolumn{1}{c}{---} & \multicolumn{1}{c}{---} \\
 $^{87}$Rb-Kr &  \multicolumn{1}{c}{---} & \multicolumn{1}{c}{---} & 2.79(1) & 2.83(4) \\
 $^{87}$Rb-Xe &  2.75(4) & \multicolumn{1}{c}{---}  & 2.88(1) & 2.93(7) \\
\hline
\end{tabular}
\caption{
Comparison of this work with published  measurements, including those utilizing universality of quantum diffractive collisions (UQDC), and theoretical calculations of $^{87}$Rb-X loss rate coefficients.
Numbers in parentheses are one-standard-deviation, $k=1$ uncertainties.
For simplicity, the statistical and systematic uncertainty from Ref.~\onlinecite{Shen2021} is added in quadrature.
For the theory and this work, $T=295.2(3)$~K. For Ref.~\onlinecite{Shen2021}, $T=294$~K.
}
\label{tab:Rb_comparison}
\end{table}

Finally, we compare our rate coefficients with those published in Refs.~\onlinecite{Klos2023,Shen2021,Shen2022} 
in Table~\ref{tab:Rb_comparison}.
Agreement is observed for $^{87}$Rb-$^4$He at the one-standard deviation ($k=1$) level.
For $^{87}$Rb-N$_2$ and $^{87}$Rb-Xe, rate coefficients based on universality of quantum diffractive collisions (UQDC) from Ref.~\onlinecite{Shen2021} are smaller than our $K$ by 12~\% and 7~\%, corresponding to more than four and two standard deviations, respectively.
The data points for $^{87}$Rb-Ar are discrepant. Further research for this system is needed.
As discussed in Ref.~\onlinecite{Shen2022} and reflected in the table, UQDC does not work well for the $^{87}$Rb-H$_2$ system.
In Ref.~\onlinecite{Shen2022}, the authors measure the ratio of loss rate coefficients for $^{87}$Rb and $^7$Li  with background H$_2$ and use the theoretical results for the $^7$Li+H$_2$ system from Ref.~\onlinecite{Makrides2019,Makrides2022Eb} to derive a loss rate coefficient for $^{87}$Rb with H$_2$. The resulting  loss rate coefficient is in agreement with Ref.~\onlinecite{Klos2023}.
This scaling procedure was first suggested by Ref.~\onlinecite{Scherschligt2017}.

\section{Conclusion} \label{sec:conclusion}

We have measured total rate coefficients for room-temperature natural abundance gas species He, Ne, N$_2$, Ar, Kr, and Xe colliding with ultracold $^7$Li and $^{87}$Rb sensor atoms using a flowmeter combined with a dynamic expansion system and two cold-atom vacuum sensors.
Our measurements have an uncertainty of better than 1.6~\% for $^7$Li and 2.7~\% for $^{87}$Rb.
We find consistency at the two-standard-deviation combined statistical and systematic ($k = 2$) uncertainty level for all gas combinations except for $^{7}$Li-Ar and $^{87}$Rb-Ar with recently published quantum-mechanical scattering calculations.\cite{Klos2023}
We also compare the rate of ``glancing'' collisions for $^{87}$Rb, collisions that do not impart enough energy to eject $^{87}$Rb from its shallow magnetic quadrupole trap, and find consistency at the two-standard-deviation combined statistical and systematic ($k = 2$) uncertainty level with the calculations of Ref.~\onlinecite{Klos2023} for all collisions except $^{87}$Rb-Kr.

An equivalent interpretation of our results is that quantum-based measurement of vacuum pressure with cold atoms is consistent with that set by a combined flowmeter-dynamic expansion standard.
Thus,  cold-atom based vacuum pressure sensors are also cold atom vacuum standards, or CAVSs.
Agreement between the dynamic expansion standard and the CAVS validates their operation as quantum-based standards for vacuum pressure.

This validation opens potential new opportunities in vacuum metrology at ultra-high vacuum (UHV) pressures.
In particular, the quantum measurement of pressure by a CAVS is {\it primary}.
It is not traceable to a measurement of like kind. 
Given the demonstrated consistency, the CAVS could now potentially replace the combined flowmeter and dynamic expansion systems in the calibration of other pressure gauges.

The portable CAVS (p-CAVS), in particular, can also replace common classical gauges, like the Bayard-Alpert ionization gauges.\cite{Eckel2018,Ehinger2022}
The p-CAVS shows lower uncertainties than calibrated ionization gauges in the UHV.\cite{Berg2015}
The performance of our p-CAVS is comparable and complementary to that of the recently developed 20SIP01 ISO ionization gauge,\cite{Jousten2021} which has better than 1.5\,\% relative uncertainties without calibration but operates at higher pressures from 10$^{-6}$~Pa to 10$^{-2}$~Pa.
Both have absolute uncertainties that are independent of the individual gauge.

Another advantage over ion gauges is related to pressure sensing with unknown mixtures of background gases.
Despite the range of masses and polarizabilities of the background gas species for which we have calculated and measured the loss rate coefficients, the maximum relative deviation of $L$ from $L$ for N$_2$ is roughly 40~\% for both $^7$Li and $^{87}$Rb, as seen in from Table~I of Ref.~\onlinecite{Klos2023}.
We believe, based on semi-classical scattering theory,\cite{Child} that the mean and variation will not significantly increase as data for other background gases become available.
Thus, we can expect a pressure measurement of (mixtures of) unknown gases by a CAVS to have at most a 40\,\% relative uncertainty if one simply used the value of $L$ for N$_2$.
The uncertainty is small compared to the factor of five difference in readings seen by an ionization gauge between N$_2$ and He at the same pressure.\cite{DushmanBook,Bartmess1983}

If the background gas contains a single species with an unknown $L$, then the procedure outlined in Ref.~\onlinecite{Booth2019, Shen2020, Shen2021} can determine $L$ from measurements at a single, unknown $n$.
The procedure relies on the validity of semi-classical scattering theory \cite{Child} and a measurement of the variation of the atom loss rate $\Gamma$ on trap depth $W$. 
The procedure is known to fail when the colliding pair's reduced mass is small compared to the cold atom's mass; the discrepancy of $K$ for $^{87}$Rb+H$_2$ between Refs.~\onlinecite{Booth2019, Shen2020, Shen2021} and Refs.~\onlinecite{Shen2022, Klos2023} is roughly 30~\%.
However, disagreements between the $K$ of Ref.~\onlinecite{Booth2019, Shen2020, Shen2021} and those of Ref.~\onlinecite{Klos2023}, mostly verified by this present work, can be between 5~\% and 9~\%, with these residual discrepancies not strongly dependent on the reduced mass.
If we ignore $^{87}$Rb+H$_2$, then, in the same spirit as the ionization gauge discussion above, we conclude that the maximum relative uncertainty for a cross section obtained using the procedure of Refs.~\onlinecite{Booth2019, Shen2020, Shen2021} is 9~\%.
Further work is required to verify the uncertainty of the methods of Ref.~\onlinecite{Booth2019, Shen2020, Shen2021}.
Because it requires knowledge of the variation of $\Gamma$ on $W$, however, the procedure will likely not be feasible for $^7$Li, given its light mass.\cite{Shen2021} 
There are simply fewer ``glancing'' collisions with which to accurately measure this dependence compared to $^{87}$Rb.

This decrease from 40~\% to 9~\% in relative uncertainty due to an unknown $L$ is not the only motivating factor in choosing between $^7$Li and $^{87}$Rb as the CAVS sensor atom.
Another key difference between $^7$Li and $^{87}$Rb is that $^{87}$Rb exhibits significant non-exponential decay in the atom-loss decay curves at the lowest UHV pressures, as evidenced by the large, fitted $\beta$ in Eq.~(\ref{eq:loss}) and shown in Fig.~\ref{fig:background_decays}.
We currently have no satisfactory explanation for this observation.
This unexpected discovery suggests that $^{87}$Rb-based CAVSs will probably not be as accurate as one based on $^7$Li in the low ultra-high vacuum and extreme high vacuum regimes.
Combined with the other advantages outlined in Ref.~\onlinecite{Eckel2018}, we believe that $^7$Li offers superior performance.

To realize the low $<2$\,\% uncertainty potential of a $^7$Li based p-CAVS, loss rate coefficients for other common gases found in vacuum chambers like CO, CO$_2$, O$_2$ and H$_2$ must be measured and compared to theoretical evaluations when available.
Measurement of $L$ with these more reactive gases requires an upgrade to our dynamic expansion system, which is currently underway.
Theoretical calculations for CO, CO$_2$ and O$_2$ are also forthcoming; theoretical calculations for H$_2$ are already contained in Ref.~\onlinecite{Klos2023}.

Finally, we must further validate the pressure range of operation of the CAVSs.
Currently, such devices have been operated as high as $6\times10^{-5}$~Pa,\cite{Shen2020} where loss rates are of the order of $10$~s$^{-1}$.
The lowest detectable pressure of a CAVS is less well characterized; we are currently endeavoring to understand the physics behind the non-exponential behavior at low pressures.

\appendix
\section{Dynamic expansion system}
\label{sec:apparatus:dynamic_expansion}

Dynamic expansion standards rely on precise knowledge of the rate of evacuation of a background gas from a vacuum chamber through an orifice.
This is achieved by using an orifice with known conductance $C_0$ that connects to a second chamber, which is evacuated using a vacuum pump with pumping speed $S$.
For $S\gg C_0$, the orifice reduces the pumping speed out of the first chamber such that the evacuation rate out of this chamber is  $C_0$, leading to
\begin{equation}
    n = \frac{\dot{N}}{C_0}\label{eq:numberdensity}\,.
\end{equation}
The flow $\dot{N}$ is both generated and measured by a flowmeter designed to operate in the XHV.\cite{Eckel2022}
The flowmeter reports a type-A, statistical $u_{\rm A}(\dot{N})$ and type-B, systematic $u_{\rm B}(\dot{N})$ uncertainty for each flow measurement. 
For this work, $u_{\rm A}(\dot{N})$ is  the larger of the extrapolated modified Allan deviation\cite{Riley2008} of $\dot{N}$ and the standard uncertainty from least-squares fitting for $\dot{N}$ from time traces of $N(t)$ in the flowmeter versus $t$.
A detailed discussion of the flowmeter is contained in Ref.~\onlinecite{Eckel2022}.

Our orifice has a cylindrical shape with a length $l = 5.0462(3)$~mm, radius $r=1.1092(4)$~cm, and a corresponding cross-sectional area $A=\pi r^2 = 3.865(3)$~cm$^2$.
The uncertainties in radius and cross sectional area are dominated by their changes along the length of the cylinder.
The orifice dimensions were obtained by NIST's dimensional metrology group using a Moore Coordinate Measurement Machine (CMM).\cite{Stoup2011}
The conductance of the orifice is given by 
\begin{equation}
C_0 = \alpha A v_{\rm th}/4\,, \label{eq:conductance}
\end{equation}
where $\alpha$ is the transmission probability of a molecule entering the orifice, and $v_{\rm th} = \sqrt{8 k T_{\rm DE}/\pi m}$ is the mean velocity in the Maxwell-Boltzmann distribution of background gas atoms or molecules with mass $m$ at temperature $T_{\rm DE}$.

For cylindrical tubes,  the transmission probability $\alpha$ is known analytically under reasonable gas flow assumptions and is only a function of $l/r$. \cite{Essen1976}
At our  uncertainties for $l$ and $r$, the transmission probability given by Eq.~(16) of Ref.~\onlinecite{Essen1976} is sufficiently accurate and gives $\alpha_{\rm An}=0.8157(1)$.
Here, the standard uncertainty $u(\alpha_{\rm An})$ follows from  uncertainty propagation of $u(l)$ and $u(r)$
ignoring correlations between the measurements of $l$ and $r$.

We amend this analytical estimate of $\alpha$ using Monte Carlo simulations of particles in our dynamic expansion standard based on the actual orifice and chamber geometries and assuming that the temperature of the particles is that of chamber walls, $T_{\rm DE}$.\cite{Kersevan:IPAC2019-TUPMP037}
In these simulations, particles only collide with the chamber walls, which is a good assumption at our UHV pressures as the mean free path for particle-particle collisions is orders of magnitude larger than the chamber sizes.
Reflections from the walls are Lambertian: the particle is given a new random  speed, sampled from the Maxwell-Boltzmann velocity distribution independent of its incoming velocity, and a random  angle $\theta$ with respect to the surface normal sampled from a $\cos\theta$ probability distribution.
Finally, particles colliding with vacuum pump surfaces have an absorption coefficient that, given the surface's area, yields the correct pumping speed.

From the Monte-Carlo simulations, we find $\alpha_{\rm MC}=0.8160(2)$, which is 0.03~\% larger than but consistent with $\alpha_{\rm An}$.
This result confirms that the chamber geometry has a negligible impact on $C_0$.
The standard uncertainty of $\alpha_{\rm MC}$ is twice that of $\alpha_{\rm An}$ as it combines two sources of (uncorrelated) uncertainty: (1) the counting uncertainty of the Monte Carlo simulations and (2) the uncertainty in the dimensions of our orifice.
We use the more conservative $\alpha_{\rm MC}$.

We measure $T_{\rm DE}$ by averaging the time-series readings of four calibrated platinum resistance thermometers (PRTs). 
The thermometers are mounted to the exterior walls of the dynamic expansion standard and are placed in pairs.
Each pair is placed on opposing sides of the standard.
One pair is coplanar with the orifice while the other pair is mounted on the first chamber 18.9(4)~cm away from the orifice plane.
A reading $T_{i,{\rm DE}}(t)$ of  thermometer $i=1$, 2, 3, or 4 at time $t$ has a standard uncertainty of 36~mK.
Self-heating of the PRTs, measured to be about 3~mK, is negligible.
Temperature gradients of approximately 0.4~K combined with drifts of roughly 0.05~K over the time interval it takes to map out the decay of sensor atom number $N_{\rm S}(t)$, however, are observed in the dynamic expansion system.
Hence, temperature gradients dominate the uncertainty of $T_{\rm DE}$ and thus $u(T_{\rm DE}) = \sqrt{s}$ with sample variance $s=\sum_{i=1}^4\sum_{j=0}^{m} (T_{i,{\rm DE}}(j\Delta t) -T_{\rm DE})^2/(4(m+1)-1)$, where time step $\Delta t=30$~s, integer $m=\lfloor t_{\rm tot}/\Delta t\rfloor$, and $t_{\rm tot}$ is the total time to acquire a measurement of a time trace $N_{\rm S}(t)$.
$T_{\rm DE}$ tracks the stabilized air temperature $T_{\rm lab}$ in the laboratory well.
For example, $T_{\rm lab} = 295.2(1)$~K and $T_{\rm DE} = 295.3(3)$~K for the data shown in Fig.~\ref{fig:background_decays}.

The temperature of the l-CAVS vacuum chamber is found by averaging the readings of four PRTs, in a manner identical to that of $T_{\rm DE}$.
Oscillations in the cooling water temperature for the electromagnets\cite{Siegel2021} that generate the l-CAVS quadrupole magnetic field causes the temperature of the l-CAVS vacuum chamber to oscillate with an amplitude of up to 0.5~K.
No temperature change is observed due to the application of current in the electromagnets.
This leads to a standard uncertainty of $u(T)= 0.3$~K for the l-CAVS, while $T$ and $T_{\rm DE}$ typically agree within their uncertainties. 

The temperature of the p-CAVS vacuum chamber is found by averaging the readings of two PRTs, in a manner identical to that of $T_{\rm DE}$.
When the p-CAVS is turned on, we empirically observe that its temperature has a time dependence  $T(t) = T_0 + \Delta T [1-\exp(-\gamma t)]$, with $T_0 \approx T_{\rm DE}$, $\Delta T\approx +5$~K, and $1/\gamma \approx 1$~h.
The temperature increase is caused by the effusive lithium source dissipating roughly 3~W of heat to evaporate lithium.
Because the outside of the p-CAVS vacuum chamber is heated above the laboratory temperature, we reasonably assume that the inside is even warmer.
Indeed, measurements with a separate, identical p-CAVS with an in-vacuum thermocouple suggest that the interior of the vacuum chamber is 1~K warmer than the exterior-mounted PRTs measure.
We conservatively take  $u(T)=|T - T_{\rm DE}|/2\approx 2.5$~K for the p-CAVS.

For the p-CAVS, we observe temperatures $T$ that significantly differ from $T_{\rm DE}$. 
That is, a temperature gradient exists between the dynamic expansion chamber and the p-CAVS and leads to `thermal transpiration'', where  equal effusive particle flux from one chamber to the other 
in the molecular-flow regime implies \cite{DushmanBook}
\begin{equation}
n = \sqrt{\frac{T_{\rm DE}}{T}} n_{\rm DE}\,,
\label{eq:correction}
\end{equation}
where $n_{\rm DE}$ is the background gas density in the dynamic expansion system and $T$ is the temperature of the background gas atoms in the CAVS.
We have also modified our Monte Carlo simulation to incorporate thermal gradients of the walls of the chambers, and find that the pressure analog of Eq.~(\ref{eq:correction}) is accurate to better than 0.4~\% assuming a temperature gradient of 10~K.

We use a turbo-molecular pump attached to the second chamber with a finite pumping speed $S\approx 1500$~L/s to evacuate the dynamic expansion system leaving a small residual pressure in this chamber and thus allowing some particles to return to the first chamber.
Equation (\ref{eq:conductance}) is derived under the assumption that particles do no return, {\it i.e.} assuming $S\to\infty$.
We can correct for the finite $S$ by measuring the pressure ratio $r_{\rm p}$ of the pressure in the first chamber to the pressure in the second chamber and using the substitution $C_0\to C_0 (r_{\rm p}-1)/r_{\rm p}$ in Eq.~(\ref{eq:conductance}).
Our measurement of $r_{\rm p}$ is described in Ref.~\onlinecite{Barker2022}. We  give a brief synopsis here.
A spinning rotor gauge (SRG) is connected via pneumatically actuated valves to either the first or the second chamber.
The SRG's decay rate, which is a proxy for the pressure, is measured sequentially as it is connected to the first and second chamber.
The ratio of these decay rates corresponds to $r_{\rm p}$.
Accurate measurements of $r_{\rm p}$ require pressures in the first chamber between 0.1~Pa and 0.6~Pa to obtain sufficient signal.
At these pressures, the non-linear conductance of the orifice needs to be accounted for and we measure pressure ratios at several pressures and linearly extrapolate to zero pressure.
The dominant uncertainty in this measurement is statistical and is typically $u(r_{\rm p})/r_{\rm p} = 0.02$.

Finally, we find that the number density of background gas at a CAVS is
\begin{equation}
    n = \sqrt{\frac{T_{\rm DE}}{T}}  \frac{\dot N}{C_0} \frac{r_{\rm p}}{r_{\rm p}-1}=   
    \frac{\dot{N}}{\alpha_{\rm MC} A}\sqrt{\frac{2\pi m}{kT}}\frac{r_{\rm p}}{r_{\rm p}-1}.
\end{equation}
by combining Eqs.~(\ref{eq:numberdensity}), (\ref{eq:conductance}), and (\ref{eq:correction}) with the substitution for $C_0$ described in the previous paragraph. 
We  use the transmission probability $\alpha_{\rm MC}$ obtained from our Monte-Carlo simulations and realize that $n$ is independent of $T_{\rm DE}$.
The  relative uncertainty $u(n)/n$  of the background gas number density at the CAVS is given by
\begin{eqnarray}
    \left[\frac{u\left(n\right)}{n}\right]^2 & = &  \left(\frac{u(\dot{N})}{\dot{N}}\right)^2 + \left(\frac{u(A)}{A}\right)^2 + \left(\frac{u(\alpha_{\rm MC})}{\alpha_{\rm MC}}\right)^2 \nonumber \\
    & & + \frac{1}{2}\left(\frac{u(T)}{T}\right)^2 + \left(\frac{1}{r_{\rm p}-1}\frac{u(r_{\rm p})}{r_{\rm p}}\right)^2 
\end{eqnarray}
assuming no correlations among the various sources of uncertainty. The contribution due to the uncertainty in $m$ is negligible for our purposes. 

Before we conclude this Appendix, let us consider the potential for pressure gradients  within the DE system at base pressure.
Differences in measured pressure at base pressure between the two CAVSs could be caused by local differences in the specific outgassing rate combined with differences of the effective vacuum conductance from each of the CAVS to the orifice.
Considering solely the latter, Monte-Carlo simulations assuming uniform specific outgassing throughout the first DE chamber and the two CAVSs  show that the l-CAVS should be at a 25~\% higher pressure than the p-CAVS because of the former's slightly longer connection to the DE chamber. 
We note that there is no guarantee that the specific outgassing of chamber walls is uniform; factors of 3-5 difference in local outgassing rates are reasonable and might explain our observations at base pressure.
Over the duration of our experiment, such imbalanced outgassing is stable.
By contrast, Monte-Carlo simulations of  the added, inert gasses, injected into the DE chamber at a specific point,
show that their partial pressure is uniform to within the simulations' uncertainty when the chamber is at uniform temperature.

\section{Imaging}
\label{sec:apparatus:imaging}

Our imaging system is a potential source of uncertainty in both the MOT atom number $N_0$ and the number of sensor atoms in the magnetic quadrupole trap $N_{\rm S}(t)$ at hold time $t$.
As described in the introduction to Sec.~\ref{sec:apparatus}, the experiment has several steps for each hold time $t$: An atom cloud is prepared in the MOT, subsequently held in the magnetic trap for a time $t$, and then atoms are recaptured into the MOT.
We take images before we load the MOT, at the moment when the MOT is fully loaded, and then after the recapture of the atoms in the MOT.
In the end, we store and analyze six images for each hold time $t$.
Specifically, before the MOT loading stage, a first image with neither the atoms nor lasers present and a second image with the MOT lasers but no atoms present are taken.
At the end of the MOT loading stage, the third image is taken and we turn off the MOT light.
These three images determine $N_0$.
This step is non-destructive.
After the recapture of the sensor atoms at time $t$, we then take three more images, spaced in time about 0.3~s apart.
The first is an image with the MOT lasers on and sensor atoms present, the second an image with the MOT lasers on and no atoms present, and finally, an image with neither laser nor atoms.
The latter three images determine $N_{\rm S}(t)$ and is destructive.

We  process or combine each set of three images using a procedure similar to that described in Appendix~A of Ref.~\onlinecite{VarennaNotes}, to account for ``dark counts'' and  differences in MOT laser intensities, and construct sensor atom number densities. We then calculate  $N_0$ or $N_{\rm S}(t)$.
For mathematical convenience, we label an image with (1) neither the atoms nor lasers present, (2) an image with the MOT lasers on and no atoms present, and  (3) an image with the MOT lasers on and sensor atoms present.

We then denote the images by $\Xi_j(\tilde{x},\tilde{y})$, where $j=1$, 2, or 3 corresponding to the image order defined in the previous paragraph, and $(\tilde{x},\tilde{y})$ correspond to the coordinates of a pixel on the camera.
The images can then be parameterized as 
\begin{eqnarray}
    \Xi_1(\tilde{x},\tilde{y}) & = & \delta(\tilde{x},\tilde{y})\,, \nonumber \\
    \Xi_2(\tilde{x},\tilde{y}) & = & \delta(\tilde{x},\tilde{y}) + q_{\rm e} G \,\Lambda(\tilde x,\tilde y) I_2\,,
    \label{eq:imaging}\\
    \Xi_3(\tilde{x},\tilde{y}) & = & \delta(\tilde{x},\tilde{y}) + q_{\rm e} G \,[\Lambda(\tilde{x},\tilde{y}) I_3 + \Omega(\tilde{x},\tilde{y}, I_3)] \,,\nonumber
\end{eqnarray}
where $\delta(\tilde x,\tilde y)$ is an image of ``dark counts'', $q_{\rm e}$ is the quantum efficiency of the camera--the probability to convert a photon into a photoelectron--and $G$ is the gain--the relationship between photoelectrons and counts on the analog-to-digital converter of the camera.
The manufacturer of our cameras specifies $G=0.072$~counts/photoelectron, $q_{\rm e}= 0.45$ for $^7$Li, and $q_e=0.30$ for $^{87}$Rb.
The function $\Lambda(\tilde x,\tilde y)$ describes how many photons are scattered from the MOT laser beams  onto pixel $(\tilde{x},\tilde{y})$ when no atoms are confined in the MOT.
Likewise, function $\Omega(\tilde x,\tilde y, I) $  describes how many photons from  atoms fluorescing in the MOT laser beams with combined or total intensity $I$ are imaged onto pixel $(\tilde{x},\tilde{y})$.
The intensities of the MOT lasers are actively stabilized, which keeps drifts and fluctuations of $I$ with time to less than $1$~\%. 
Nevertheless, we  correct for residual changes of laser intensities $I_j$ with $j=2$ and 3.

The dimensionless function $\Omega$ is given by 
\begin{eqnarray}
    \Omega(\tilde{x},\tilde{y}, I) &=& \frac{1-\sqrt{1-{\rm NA}^2}}{2}
    \left(\frac{\Delta \tilde{x}}{M}\right)^2 t_{\rm e} \label{eq:Omega} \\
    && \times \int_{-\infty}^{\infty} {\rm d}z\ 
      n_{\rm S}(\tilde x/M,\tilde y/M,z)
 \,  R(\tilde x/M,\tilde y/M,z, I)
    \,,  \nonumber
\end{eqnarray}
where the dimensionless ${\rm NA}$ and $M$ are the numerical aperture and magnification of the imaging system, respectively.
The quantity $\Delta \tilde{x}$ is the length of a side of the square pixels in the camera, $n_{\rm S}(\vec x)$  is the number density of sensor atoms at position $\vec x=(x,y,z)$ in the MOT, $R(\vec x,I)$ is a position and intensity-dependent scattering rate in the MOT, and $t_{\rm e}$ is the exposure time of the camera.
Equation (\ref{eq:Omega}) is valid when magnification $M$ does not vary over the size of the MOT and  the depth of field is larger than size of the MOT, both reasonable approximations for our imaging system.
It also assumes that the atoms fluoresce equally into  $4\pi$ sterradians.

A determination of $\Omega(\tilde{x},\tilde{y}, I_3)$ is required to obtain $N_S$ and $N_0$.
We manually define a region of interest (ROI) that includes the region where sensor atoms are located in image $j=3$.
The size of the ROI is less than 20\,\% of the total image size.
The ratio
\begin{equation}
    r_{\rm I} \equiv  \frac{\sum_{(\tilde x,\tilde y) \notin {\rm ROI}}\, [\Xi_2(\tilde{x},\tilde{y})-\Xi_1(\tilde{x},\tilde{y})]}{\sum_{(\tilde{x},\tilde{y}) \notin {\rm ROI}} \,[\Xi_3(\tilde{x},\tilde{y})-\Xi_1(\tilde{x},\tilde{y})]}  = \frac{I_2}{I_3}\,,
\end{equation}
where the sums are over all pixels outside the ROI, is then equal to the ratio of laser intensities used for images $j=2$ and 3.
Next, we realize that
\begin{eqnarray}
  \pazocal{W}_3(\tilde{x},\tilde{y})&\equiv&
    \Xi_3(\tilde{x},\tilde{y}) - \Xi_1(\tilde{x},\tilde{y}) -
    \frac{1}{r_{\rm I}} [\Xi_2(\tilde{x},\tilde{y})-\Xi_1(\tilde{x},\tilde{y})]  \nonumber\\
    &=& q_{\rm e} G\Omega(\tilde{x},\tilde{y},I_3) \,.
\end{eqnarray}
We have verified that this reconstruction of $\pazocal{W}_3(\tilde{x},\tilde{y})$ and thus $\Omega(\tilde{x},\tilde{y},I_3)$ yields
\begin{equation}
\sum_{(\tilde x,\tilde y) \in {\rm ROI}} \pazocal{W}_3(\tilde{x},\tilde{y}) = 0
\end{equation}
when $n_{\rm S}(\vec x)=0$ for all $\vec x$.

To obtain $N_{\rm S}$ or $N_0$ from $\pazocal{W}_3(\tilde{x},\tilde{y})$,  we use the approximation that the scattering rate $R(\vec x, I)$ is independent of $\vec x$
and given by
\begin{equation}
   R(\vec x, I) = \frac{1}{2\tau} \frac{I/I_{\rm sat}}{1 + I/I_{\rm sat} + 4(\tau \Delta )^2}
   \equiv R_0(I)\,,
   \label{eq:Rapprox}
\end{equation}
where $I_{\rm sat}$ is the two-level saturation intensity of the atomic cycling transition, $\tau$ is the excited state lifetime, and frequency $\Delta$ is the laser detuning from the atomic transition.
For our MOTs, we operate at $\tau\Delta = -2$.
The detuning $\Delta$ exhibits short-term relative fluctuations of $<4$~\% with no detectable long-term drifts.
The MOTs operate in the non-saturated regime where $R(\vec x, I) \propto I$.
In addition, to eliminate systematic effects from changes of the two $I_j$ with time $t$, we also compute the quantity
\begin{equation}
 \pazocal{L}_3\equiv
     \sum_{(\tilde{x},\tilde{y}) \in {\rm ROI}} \frac{1}{r_{\rm I}}  [\Xi_2(\tilde{x},\tilde{y})-\Xi_1(\tilde{x},\tilde{y})]= q_{\rm e} G I_3   \sum_{(\tilde{x},\tilde{y}) \in {\rm ROI}}\Lambda(\tilde{x},\tilde{y}) \,.
\end{equation}
The sensor atom number is finally given by
\begin{equation}
    N_i = \frac{2}{1-\sqrt{1-{\rm NA}^2}}\frac{1}{q_e G t_{\rm e}}\frac{\langle \pazocal{L}_3 \rangle}{R_0(\langle I_3\rangle)\pazocal{L}_3}
  \sum_{ (\tilde{x},\tilde{y})\in {\rm ROI} }\pazocal{W}_3(\tilde{x},\tilde{y})\,,
\end{equation}
where $\langle \pazocal{L}_3\rangle$ is the average value of $\pazocal{L}_3$ over the multiple repetitions of the experiment measuring $N_0$ or $N_{\rm S}(t)$ for the same time $t$.
Here, forming ratio $\langle\pazocal{L}_3\rangle/[R_0(\langle I_3\rangle)\pazocal{L}_3]$ eliminates fluctuations of the scattering rate due to laser fluctuations about its time-averaged value of $\langle I_3\rangle$, which is independently measured with a power meter and the known $1/e^2$ MOT beam radius.
This procedure eliminates any potential correlations between $I_3$ and $t$.

Finally, the ratio 
\begin{equation}
\eta_{\rm S}(t) = \frac{N_{\rm S}(t)}{N_0} \label{eq:etaS}
\end{equation} 
is formed from the independently measured $N_0$ and $N_{\rm S}(t)$. 
This ratio eliminates the effect of the uncertainties in NA, $q_{\rm e}$, $G$, and $t_{\rm e}$.
As described in Sec.~\ref{sec:loss_curves}, we observe $u(\eta_{\rm S}(t))/\eta_{\rm S}(t)<0.03$ for the p-CAVS and $u(\eta_{\rm S}(t))/\eta_{\rm S}(t)<0.05$ for the l-CAVS for any single measurement at short time $t$.
This statistical uncertainty is most likely due to short-term fluctuations in $\tau\Delta$ and fluctuations in the fraction of atoms successfully transferred from the MOT to the magnetic trap.
At long $t$, the fluctuations are determined by the statistical noise in the camera and reflect a minimum detectable atom number.

We last consider correlations between sensor atom number density $n_{\rm S}(\vec x)$ and $t$, or, equivalently, correlations between the shape of $n_{\rm S}(\vec x)$  and  $N_{\rm S}(t)$.
Most easily inferred from Eq.~(\ref{eq:Omega}), the sensor atom number is proportional to a three-dimensional integral with an integrand that is the product of $n_{\rm S}(\vec x)$ and scattering rate $R(\vec x, I)$.
The spatial dependence of $R(\vec x, I)$ can be found by generalizing Eq.~(\ref{eq:Rapprox}).
We include spatially-dependent Zeeman shifts in the detuning $\Delta$ and a spatially dependent laser intensity. Combined with the variation of the shape of $n_{\rm S}(\vec x)$ with $N_{\rm S}$, this produces a systematic relative uncertainty in our calculated $\eta_{\rm S}(t)$ of $<3$~\%.
This ``imaging stability'' uncertainty is propagated through the fitting described in Secs.~\ref{sec:loss_curves} and \ref{sec:analysis}.

We note that the use of subtracted images assumes linearity between the number of photons incident on the camera and the number recorded by the 10-bit analog-to-digital converter of the camera.
CMOS cameras, in particular, are known to be non-linear, with most of the non-linearity coming from the amplification system.
We have independently measured the non-linearity of our cameras and analyzed our results with and without accounting for the camera non-linearity, and found a relative uncertainty correction to $\Gamma$ of only 0.07~\% on average, which we take as a $k=1$ a systematic uncertainty.

Finally, our analysis also assumes linearity between the number of fluorescence photons and number of atoms in the MOT.
For optically thick MOTs, the input beams are attenuated, leading to less overall fluorescence.
For $N\sim 10^5$, the p- and l- CAVS MOTs have 0.1 and 0.3 peak resonant optical depth, respectively, leading to an attenuation of the detuned MOT beams as they traverse the atomic cloud of  0.1~\% and 0.3~\%, respectively.
This attenuation causes a slight undercount of atoms at early times.
When fitting time traces of $\eta_{\rm S}$ with Eq.~(\ref{eq:loss}), this effect manifests predominantly as a negative value for $\beta$, which we do not observe in our experimental data.
Simulations with noiseless data show that relative shift in $\Gamma$ is at a negligible  $10^{-6}$ level.

\section*{Acknowledgements}
The authors thank L. Ehinger, P. Elgee, and A. Sitaram for initial development of the p-CAVS; B. Acharya, E. Newsome, and R. Vest for technical assistance; N. Klimov for fabrication of the grating-MOT chip; E. Norrgard and W. Phillips for useful discussions; and K. Douglass and G. Fraser for a thorough reading of the manuscript.

\section*{Author Declarations}
\subsection*{Conflicts of Interest}
D.S.B., J.A.F., J.S., and S.P.E. have U.S. patent 11,291,103 issued. D.S.B. and S.P.E. have U.S. provisional patent 63/338,047 filed.

\section*{Data Availability}
The data that support the findings of this study are available from the corresponding author upon reasonable request.

\bibliography{main}

\end{document}